\documentclass[prd,nofootinbib]{revtex4}
\usepackage{amsmath}
\usepackage{amssymb}
\usepackage{graphicx}
\usepackage[usenames, dvipsnames]{color}

\newcommand{\tpk}{T_{\rm peak}}
\newcommand{\tx}{T_{\rm cross}}
\newcommand{\omax}{\omega_{\rm max}}

\begin{document}

\title{Arrival Times of Gravitational Radiation Peaks for Binary Inspiral}

\author{Richard H.~Price} \affiliation{Department of Physics, MIT, 77 Massachusetts Ave., Cambridge, MA 02139}
\affiliation{Department of Physics, University  of Massachusetts, Dartmouth, MA 02747}

\author{Gaurav Khanna} 
\affiliation{Department of Physics, University
  of Massachusetts, Dartmouth, MA 02747}

\begin{abstract}
Modeling of gravitational waves (GWs) from binary black hole inspiral
brings together early post-Newtonian waveforms and late quasinormal
ringing waveforms.  Attempts to bridge the two limits without recourse
to numerical relativity involve predicting the time of the peak GW
amplitude. This prediction will require solving the question of why
the peak of the ``source,'' i.e., the peak of the binary angular
velocity, does not correspond to the peak of the GW amplitude.  We
show here that this offset can be understood as due to the existence
of two distinct components of the radiation: the ``direct'' radiation
analogous to that in flat spacetime, and ``scattered'' radiation
associated with curved spacetime.  The time dependence of these two
components, and of their relative phases determines the location of
the peak amplitude.  We use a highly simplified model to clarify the
  two-component nature of the source, then demonstrate that the
  explanation is valid also for an extreme mass ratio binary inspiral.
\end{abstract}

\maketitle

\section{Overview}\label{sec:overview}
\subsection{Introduction and Summary}

\medskip
The recent detections of gravitational radiation events GW150914~\cite{GWPRL1}
and GW151226~\cite{GWPRL2} from black hole binary inspiral
has underscored the importance of understanding the
theoretical predictions of the waveforms for such processes. Numerical
relativity, the computation of the Einstein's nonlinear
equations~\cite{NR}, has played an important role in generating these
predictions, but there is a wide range of binary parameters, and
studying an appropriately large and finely spaced set of waveforms
requires a more efficient methodology to supplement numerical
relativity. One such methodology is the effective-one-body
approximation (EOB)~\cite{EOB1,EOB2,BB}, used together with the linear
equations of the extreme mass ratio particle-perturbation
technique~\cite{BBHKOP,TBKH} and numerical relativity~\cite{PBBBKPS}.

In the EOB model, two different techniques are used to generate a full waveform  
signal from a black hole binary system and the results are amplitude 
and phase matched. In the early epoch (inspiral-plunge) of the binary's 
evolution, an EOB Hamiltonian system is evolved and the waveform is computed 
using an inspiral-plunge trajectory. For the late (merger-ringdown) epoch, 
the early epoch waveform is matched to a linear superposition of several 
quasi-normal modes. The precise moment in the binary's evolution at which this matching is 
performed is thus a very critical matter in the EOB approach. In the original 
EOB work~\cite{EOB1,EOB2} it was argued that this matching should be performed 
close to the light ring where the orbital frequency peaks; thus, the peak of the 
waveform amplitude was identified with the light ring as well. 

The fact that these two peaks do not actually align has been a complication 
in this program. Particularly troublesome has been the dependence of the peak
offsets on the orbital frequency, and the fact that different
(spin-weighted) spherical harmonic modes have significantly different
offsets~\cite{BBHKOP,TBKH,PBBBKPS}.  In the EOB papers a time delay is
defined, $\Delta t^{\ell,m}_{\rm peak}=t^{\ell,m}_{\rm peak}
-t^{\Omega}_{\rm peak} $, where $t^{\ell,m}_{\rm peak}$ is the time at
which the peak of the ${\ell, m}$ mode of radiation is observed, and
$t^{\Omega}_{\rm peak}$ is the time at which radiation would be
observed from the maximum of the particle's angular velocity. In most
cases (i.e. results from numerical relativity and particle perturbation 
theory models), this delay is negative; the peak radiation is earlier than
the radiation from the peak of angular velocity.

Such studies have raised the question why there should be a peak offset,
that is, why the apparent peak of the supposed source of radiation does not
align with the peak of the radiation generated. Not only would an
answer satisfy scientific curiosity, but it could lead to an effective
way of predicting the peak offsets. It has been
suggested~\cite{ScottField} that the offset, and its dependence on
angular frequency, might be explained as manifestations of the energy
dependence of the scattering of gravitational radiation off the
curvature potential~\cite{curvpot} created by the black holes.

In this paper we will answer that question by exploiting an approach
we have recently developed~\cite{PNK}. In that approach, we use the
Fourier domain Green function (FDGF) for a very simple model in which
we replace the curvature potential with a simplified truncated dipole
potential (TDP)~\cite{tdp}. In our approach, we use point particle
trajectories, but we do not require them to be analogs of geodesics.
This greater freedom to vary trajectories, combined with the 
simplification from the TDP model, gives transparency to the problem of
the location of the radiation peak, and its dependence on angular
velocity.

The clear picture that is revealed is that the outgoing radiation
consists of two components. One component, the quasinormal or
scattered radiation, is due to the motion through the strong field
spacetime (equivalently, motion in the neighborhood of the peak of the
curvature potential). The second component is the ``direct
radiation,'' radiation that has little to do with the curved
spacetime, and can be fully ascribed to the motion of the particle as
if it were in flat spacetime.  The peak radiation is a result of the combination of
the two components.

Through the use of numerical evolution codes we have investigated
whether this picture applies to the extreme mass-ratio inspiral (EMRI)
problem in the Schwarzschild geometry. Although the direct and scattered
contributions cannot be separated in an evolution code, the character of 
the results clearly indicates that this picture does apply. Further study
is appropriate to the applicability to the inspiral of comparable mass holes,
but the indications are strong that here too it applies.

The remainder of this paper is organized as follows. In
Sec.~\ref{sec:TDP} we review the features of the TDP/FDGF model that
are most relevant to the investigation of the peak location. Numerical
results from that model are presented in Sec.~\ref{sec:TDPresults},
and a tentative explanation is presented of the location of the
amplitude peak. Section~\ref{sec:Schw} then shows that the qualitative
features of the peak location are the same for the EMRI models in the
Schwarzschild spacetime as in the TDP model. Conclusions are
given in Sec.~\ref{sec:conc}.

\section{The TDP/FDGF Methodology}\label{sec:TDP}

When analyzed into spherical harmonics of index ($\ell,m$) perturbations
of Schwarzschild black holes can be formulated in terms of a radiation
field (scalar, electromagnetic, gravitational,\dots) satisfying an
equation of the form
\begin{equation}\label{eq:Schweq}
  \frac{\partial^2\Psi_{\ell m}}{\partial r^{*2}}
-\frac{\partial^2\Psi_{\ell m}}{\partial t^2}
-V_{\ell}(r^*)\Psi_{\ell m}=S_{\ell m}(r^*,t)\,.
\end{equation}
Here $r^*$ is the Regge-Wheeler~\cite{RW57} ``tortoise coordinate''
which places the event horizon at $r^*=-\infty$, and approaches the ordinary
Schwarzschild areal radius $r$  when  $r^*$ is much larger than the mass $M$.
The details of the curvature potential $V_{\ell}(r^*)$ depend on which perturbation field is being
represented, but in all cases $V_{\ell}(r^*)\rightarrow \ell(\ell+1)/r*^2$
as $r^*/M\rightarrow\infty$, and $V_{\ell}(r^*)\rightarrow \mbox{const}\times
e^{r^*/2M}$ as $r^*/M\rightarrow-\infty$.
In the TDP
model of a scalar field $\Psi$, the radial variable is replaced by a variable $x$  that ranges from 
$-\infty$ to $+\infty$, and the $\ell=1$ curvature potential is replaced by
\begin{equation}\label{eq:TDPpot}
  V=\left\{
    \begin{array}{cl}
    \ell(\ell+1)/x^2=  2/x^2&{\rm for}\ x>x_0\\0&{\rm for}\ x<x_0 
    \end{array}
\right.\,.
\end{equation}
The potential ``edge,'' or peak, at $x_0$, is the TDP proxy for the
peak of the curvature potential approximately at the $r=3M$ location
of the circular photon orbit, the ``light ring'' (LR), in the
Schwarzschild geometry.

A scalar field $\Psi$ is imagined to have a point particle source, so
that the time domain equation for $\Psi$ is
\begin{equation}\label{eq:xeq}
\frac{\partial^2\Psi}{\partial x^{2}}
-\frac{\partial^2\Psi}{\partial t^2}
-V(x)\Psi=f(t)\delta\left(x-F(t)\right)\,.
\end{equation}
In the source, the function $F(t)$ represents the position  of 
the point particle as a function of time. The function $f(t)$ can represent
a time-dependent modulation of the source; we shall use it below 
to account for the effects of angular motion. The Fourier domain Green 
function ${\cal G}$ for a particle at $x=\widetilde{F}$ satisfies
\begin{equation}\label{eq:FDGF}
 \frac{\partial^2{\cal G}}{\partial x^2}+\left(\omega^2-V(x)\right) 
{\cal G}=\delta(x-\widetilde{F}),
\end{equation}
and with it the solution to Eq.~(\ref{eq:xeq}) takes the form
\begin{equation}\label{eq:PsifromcalG}
  \Psi(t,x)=\frac{1}{2\pi}\int_{-\infty}^\infty\int_{-\infty}^\infty\int_{-\infty}^\infty
  e^{-i\omega(t-T)} {\cal G}(x,\widetilde{F};\omega)
  f(T)\delta(\widetilde{F}-F[T])\,d\omega\,d\widetilde{F}\,dT\,.
\end{equation}

\subsection{Model details}
In particle-perturbation studies, and in our TDP model, there is no
constraint that the particle trajectory be associated with geodesic
motion, so that we are free to choose families of trajectories that 
best probe the questions of interest.
For the ``radial,'' i.e., ``$x$'' motion~\cite{radial}, we choose
$F(t)$
to be a relatively simple algebraic function
\begin{equation}\label{eq:FofT}
  F(t)=\left\{
  \begin{array}{cl}
    a_0+\tau-\left(t^3+\tau^3\right)^{1/3}&t>0\\a_0&t<0\,.
  \end{array}
  \right.       
\end{equation}
The two parameters are $a_0$, the initial $x$ position of the
particle, and $\tau$, the timescale on which the particle
accelerates. Note that the particle starts, at $t=0$, with 
zero velocity and acceleration, so that the source can be considered
to have been stationary prior to $t=0$.

If the TDP model is an analog of a particle with azimuthal position
$\phi=\widetilde{\phi}(t)$ we take $f(T)=e^{i\widetilde{\phi}(T)}$\,.
The reason behind this choice is as follows: The function $\Psi$ in
Eq.~(\ref{eq:xeq}) is actually the analog of the coefficient function
$\Psi_{\ell m}(t,r)$ of a spherical harmonic expansion, confined to
$\ell=1$. In the case of radial motion only the $m=0$ component is
nonzero (with the radial motion taken along the $\theta=0$ axis of the
spherical coordinates).  In the case of angular motion, $\theta$ is
taken as $\pi/2$, and only the $m=\pm1$ components are nonzero. The
choice $f(T)=e^{\mp i\widetilde{\phi}(T)}$ corresponds to $m=\pm1$.

In order to represent a particle that is stationary in angular as well as 
radial position prior to the $t=0$ start of motion, we choose as our model
\begin{subequations}\label{eq:phiofT}
\begin{align} 
 \tilde{\phi}(t)&=\omega_{\rm max}T_{\rm peak}\,
\frac{e^3}{81}\left[e^{-z}(-6-6z-3z^2-z^3)+6\right]\\
z&\equiv 3 t/T_{\rm peak}\,.
\end{align}\end{subequations}
For this choice, the angular velocity is 
\begin{equation}
  \omega\equiv\frac{d\tilde{\phi}}{dt}=\omax\,\frac{e^3}{27}
\left(z^3e^{-z}\right)\,.
\end{equation}
The angle $\tilde{\phi}$, and the angular velocity and angular acceleration, are zero
at $t=0$, so that for its angular motion as well as its ``radial'' motion, the particle can be taken to have been stationary prior to $t=0$.
Of particular importance is the fact that we can independently specify the parameter
$\omega_{\rm max}$, the maximum angular velocity, and 
$T_{\rm peak}$, the particle time at which that maximum is achieved.

\subsection{Closed form solution}
For the TDP model, the FDGF can be found in closed form for each of
the relative positions of the potential ``edge,'' $x_0$, the particle
location $\widetilde{F}$, and the location $x$ at which the field is
being calculated. Since we are interested in the radiation region, we
consider $x$ to be unbounded. We are left then with the case in which
$\widetilde{F}>x_0$, the particle, in its inward motion has not yet
crossed the edge at $x_0$, and $\widetilde{F}<x_0$, the case in which
the particle has already passed inside the edge.  In our numerical
studies we have found that the early retarded time peak of the outgoing radiation, the 
peak of interest as an analog of the Schwarzschild problem, is
 generated by the motion of the particle before it
passes inward past the edge.  In this case, the radiative part of the
solution (i.e., the part to leading order in $1/x$) has the form
\begin{equation}\label{eq:psigen}
  \psi(t,x)=
\frac{1}{2}\,\int_{T_2}^{T_1}
\frac{u-T}{F(T)}
f(T)\,dT
    -\,\frac{1}{2}\,\int_{-\infty}^{T_2}
\,e^{-\gamma_I}\left[
-(\cos{\gamma_R}+\sin{\gamma_R})+\frac{2x_0}{F(T)}\,\cos{\gamma_R}
\right]f(T)\,dT\,.
\end{equation}
Note that this solution is good only for retarded time $u\equiv t-x$ less than or equal to the 
critical value  $u_{\rm cross}\equiv T_{\rm cross}-x_0$, where $T_{\rm cross}$  
is the time at which the particle reaches $x_0$.  

In the solution given by Eq.~(\ref{eq:psigen}) the
$t,x$ dependence is contained in the dependence 
on retarded time $u\equiv t-x$
of the integration limits $T_1, T_2$ and of the real and imaginary 
parts of $\gamma$, given by 
\begin{equation}\label{eq:thetgamdef}
 \gamma_R=\omega_R\left(u-T-F(T)+2x_0\right)
\quad\quad
 \gamma_I=\omega_I\left(u-T-F(T)+2x_0\right)\,.
\end{equation}
Here $\omega_R$ and $\omega_I$ are the real and imaginary part
of the TDP quasinormal frequency. For the TDP, there is only a single
QN frequency, a frequency which turns out to have real and imaginary parts of equal
size
\begin{equation}\label{eq:wqn}
  \omega_R=\omega_I=1/(2x_0)\,.
\end{equation}

\begin{figure}[h]%T1T2 illustration
  \begin{center}
  \includegraphics[width=.25\textwidth ]{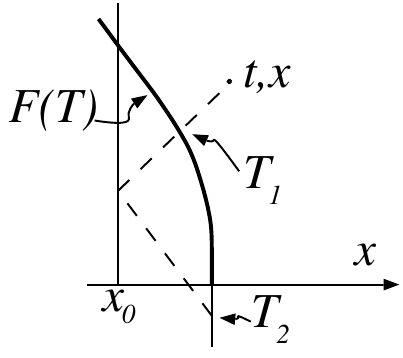}
  \caption{ A spacetime diagram showing the meaning of $T_1$ and
    $T_2$. The bold line represents the trajectory of the particle
    $x=F(T)$; the dashed lines are radial characteristics.  }
  \label{fig:T1T2}
  \end{center}
  \end{figure}

The notation and meaning of $T_1$ and $T_2$ are taken from our paper,
Ref.~\cite{PNK}, and are defined by
\begin{equation}\label{eq:T1T2}
  T_1=u+F(T_1)\quad\quad T_2=u+2x_0-F(T_2)\,,
\end{equation}
and are illustrated by the spacetime cartoon in Fig.~\ref{fig:T1T2}
from our earlier paper~\cite{PNK}.  This figure, and
Eq.~(\ref{eq:T1T2})
show that $T_1$ is simply the causality limit; the observation event
cannot be influenced by a point on the trajectory at a later retarded
time. 
The meaning of the $T_2$ limit is more subtle; it restricts the
second integral to times less than those for which the particle source
had time to ``bounce'' its influence off the potential edge and reach
the observation point $t,x$.  The $T_2$ upper limit, and the presence
of $\gamma_R,\gamma_I$ in the second integral are strong signs that
the second  integral represents the scattering of the particle influence from
the potential.

The first integral has a very different interpretation. The integrand
suggests that this integral represents ``direct'' radiation, with no
reference to the quasinormal modes or to potential scattering~\cite{explainT2}.
We can make this connection much more convincing by considering the direct scalar 
radiation generated by a particle in flat spacetime. This result is
\begin{equation}
  \psi_{\rm flat}(t,x)=
\frac{1}{2}\,\int_{\bar{T}_2}^{\bar{T}_1}
\frac{u-T}{F(T)}
f(T)\,dT\,,
\end{equation}
where
\begin{equation}\label{eq:T1T2flat}
  \bar{T}_1=u+F(\bar{T}_1)\quad\quad \bar{T}_2=u-F(\bar{T}_2)\,.
\end{equation}
The first integral in Eq.~(\ref{eq:psigen}), rather clearly,
represents direct radiation with the $x_0$ edge replacing the
coordinate origin as the location at which ingoing radiation is
reversed and sent outward. We shall, henceforth, refer to the first
integral in Eq.~(\ref{eq:psigen}) as the direct contribution to the
radiation and shall refer to the second integral as the scattered, or QN
contribution.

\section{The TDGF/TDP Results}\label{sec:TDPresults}

With $f(T)=\exp{(-i\tilde{\phi}(T))}$, Eq.~(\ref{eq:psigen}) gives us
$\psi_{1,1}(t,x)$, the $\ell, m=1,1$ spherical harmonic part of the
radiation; with $f(T)=\exp{(+i\tilde{\phi}(T))}$, it gives us the
$\ell, m=1,-1$ part. We present results in this section for the
amplitude of the radiation
\begin{equation}
\psi_{\rm amp}=|\psi_{1,1}|=|\psi_{1,-1}|\,.
\end{equation}
These results were also computed with the evolution code described in Sec.~\ref{sec:Schw}, 
in which the Schwarzschild curvature potential was replaced by the TDP potential. This check
showed accurate agreement with the results of the FDGF integration.

We shall focus on a particular radial motion, that for the following
trajectory parameters in Eq.~(\ref{eq:FofT}): $a_0=4$, 
$\tau=10$, $x_0=1$~\cite{scaling}. For these parameters, the particle, on its
inward journey, crosses the $x_0$ edge at $T_{\rm cross}=10.6177$,
with a rather relativistic speed $v_{\rm cross}=|dF/dT|_{\rm
  cross}=0.6671$.  
We have checked other ``radial'' motions, including trajectories with much 
smaller values of $v_{\rm cross}$, to be sure that the observations 
in this paper apply rather generally.

There are two important considerations in the choice of the parameters
for the angular motion in Eq.~(\ref{eq:phiofT}). First, for the
Schwarzschild EMRI case the maximum of the angular velocity is at the
$r=3M$ light ring. The TDP analog of the light ring is the potential
edge at $x_0$.  The problem is that this is also where the maximum
radiation would be expected due to features of the problem that have
nothing to do with angular motion. (This expectation will be confirmed
by results below.) Thus, if we use the most physical choice for $\tpk$, the choice
$\tpk=\tx$, it will be impossible to disentangle the effect of angular motion from
other elements of the generation of radiation. For that reason we choose 
$\tpk=\tx+\mbox{shift}$, and make three choices of the value of the shift parameter.

The second important consideration is the scale for
$\omax$. Intuitively we (correctly) suspect that too small a value of
$\omax$ will mean that the angular motion will be irrelevant; the
properties of the radiation will differ very little from those for radial
infall. We suggest that angular motion will become significant when
$\omax$ is of the order of the quasinormal frequency $1/2x_0$. A justifiable  
basis for such intuition is that the only length scale in the model is $x_0$, so 
the only scale for frequency is $1/x_0$. Full clarification will come later in 
this section. 
\begin{figure}[h]
  \begin{center}
  \includegraphics[width=.5\textwidth ]{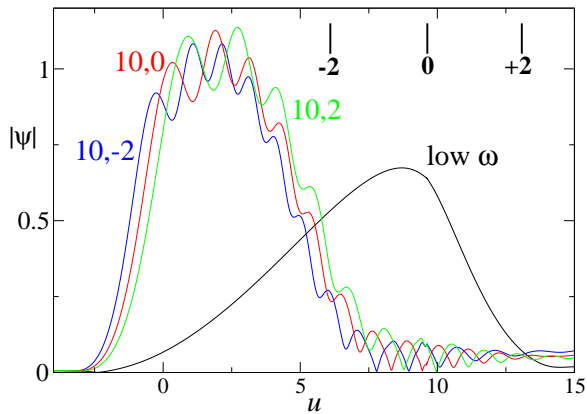}
  \caption{Amplitude as a function of retarded time $u\equiv t-x$ for
    the standard model of ``radial'' infall with $a_0=4$, $\tau=10$,
    $x_0=1$. The curve marked ``low $\omega$'' represents all models
    with $\omax\leq0.1$. The numerical markings on the higher curves
    indicate the value of $\omax$ and the shift, respectively. The
    short vertical bars indicate the retarded time at which a signal
    would be received from the particle when moves at $\omax$.  The
    bars are marked with the value of the applicable
    shift.\label{extremes}}
%this is the caption   9.6177, 6.41045, 13.0544
  \end{center}
\end{figure}

We first dispense with the high and low cases in
Fig.~\ref{extremes}. For $\omax\leq0.1$, all curves, for any of the
shifts, are indistinguishable, and are represented in
Fig.~\ref{extremes} by the single curve denoted ``low $\omega$.'' This
simply confirms the understanding that for such values the angular
motion is irrelevant to the generation of the peak radiation. The peak
of the amplitude of this curve is at $u=8.7$, reasonably close to the
retarded time $u_{\rm cross}=\tx-x_0=9.6177$ corresponding to the
particle crossing the potential edge.  This is also the time when the
particle achieves its maximum angular velocity for the choice
shift=0. The choices shift=$\pm2$, however, correspond to very
different retarded times, as indicated by the vertical bars in the
figure. The fact that the low $\omega$ curves are not influenced by a
change in the shift choice, emphasizes that these results show no
effects of angular motion.

The higher curves in Fig.~\ref{extremes} show the radiation for
$\omax=5$, and for shifts 0,$\pm2$ (given as the second index in each
curve's label).  These curves do not all have a well defined peak, but
a few points remain quite clear: (i)~The location of the region of
peak amplitude is significantly earlier than the time that peak
radiation would be expected (the vertical bars in the figure) if the
maximum amplitude were related to the maximum angular
velocity. (ii)~The displacement of the amplitude peaks has no strong
correlation with the shifts in the peak angular velocity.  (iii)~The
curves show an oscillatory behavior throughout that can be shown to
have periodicity that is accurately correlated with the particle
angular motion at the corresponding retarded time, after a correction
for the doppler shift~\cite{PNK}. (iv)~The fact that there are
oscillations shows that the radiation cannot be due only to the
angular motion.  (The $e^{i\tilde{\phi}(t)}$ source motion has a
constant magnitude.) The oscillations are proof that the total
radiation is a mixture of contributions associated with the angular
motion and those that are not.

The qualitative nature of the results in Fig.~\ref{extremes} for
$\omax=5$ are representative of all high $\omega$ models, all models
with $\omax$ greater or equal to around 4. At higher $\omax$, the
curves have a higher frequency oscillation, but are located in the
same general range of retarded time.

\begin{figure}[h]
  \begin{center}
  \includegraphics[width=.7\textwidth ]{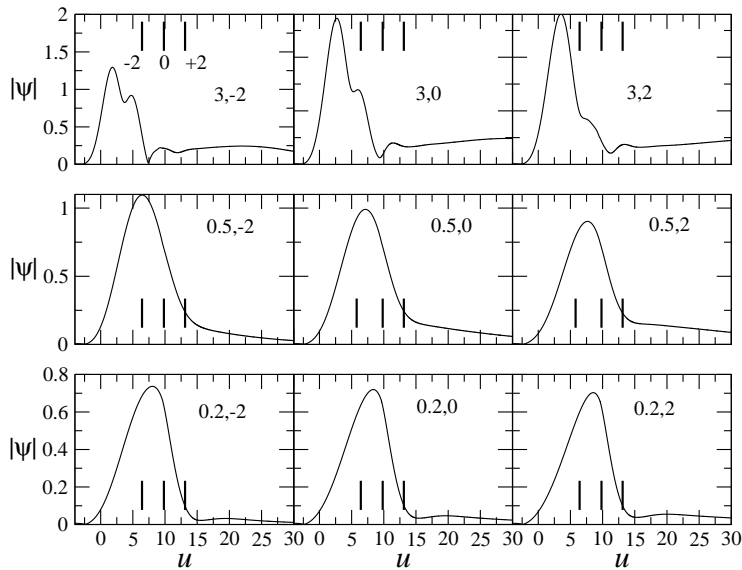}
  \caption{Amplitude as a function of retarded time for the
standard model, as in Fig.~\ref{extremes}, but here for
intermediate values of $\omax$.  As in Fig.~\ref{extremes},
curves are labeled with indices of $\omax$ and shift. The bold
vertical bars represent the retarded times corresponding to the
peak of $\omega$, for shifts of $0,\pm2$, as labeled in the first
graph. Note that the QN frequency's value is $0.5$ for this
model.  
\label{intermediate}}
  \end{center}
\end{figure}

\begin{figure}
  \begin{center}
     \includegraphics[width=.45\textwidth ]{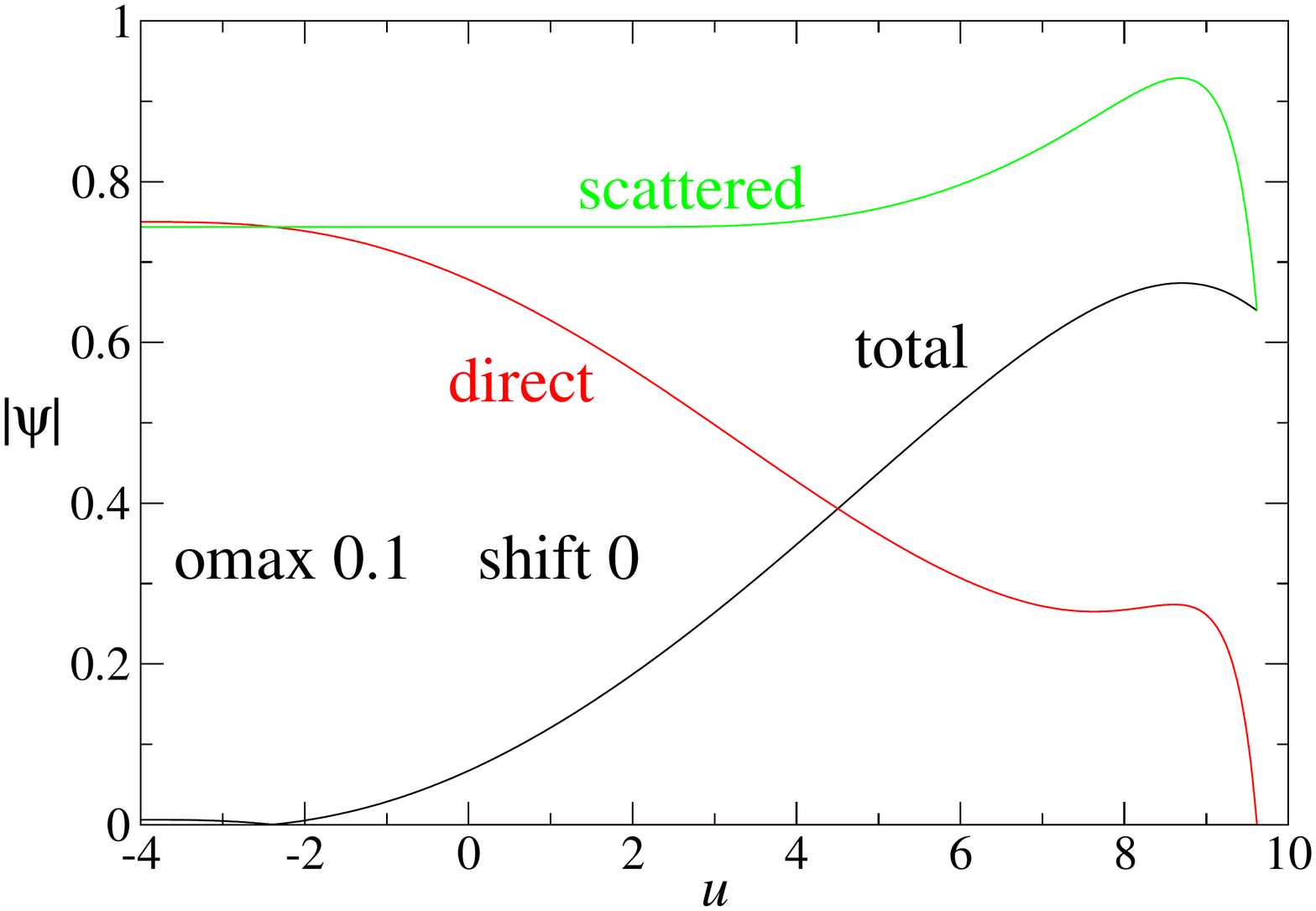}
     \includegraphics[width=.45\textwidth ]{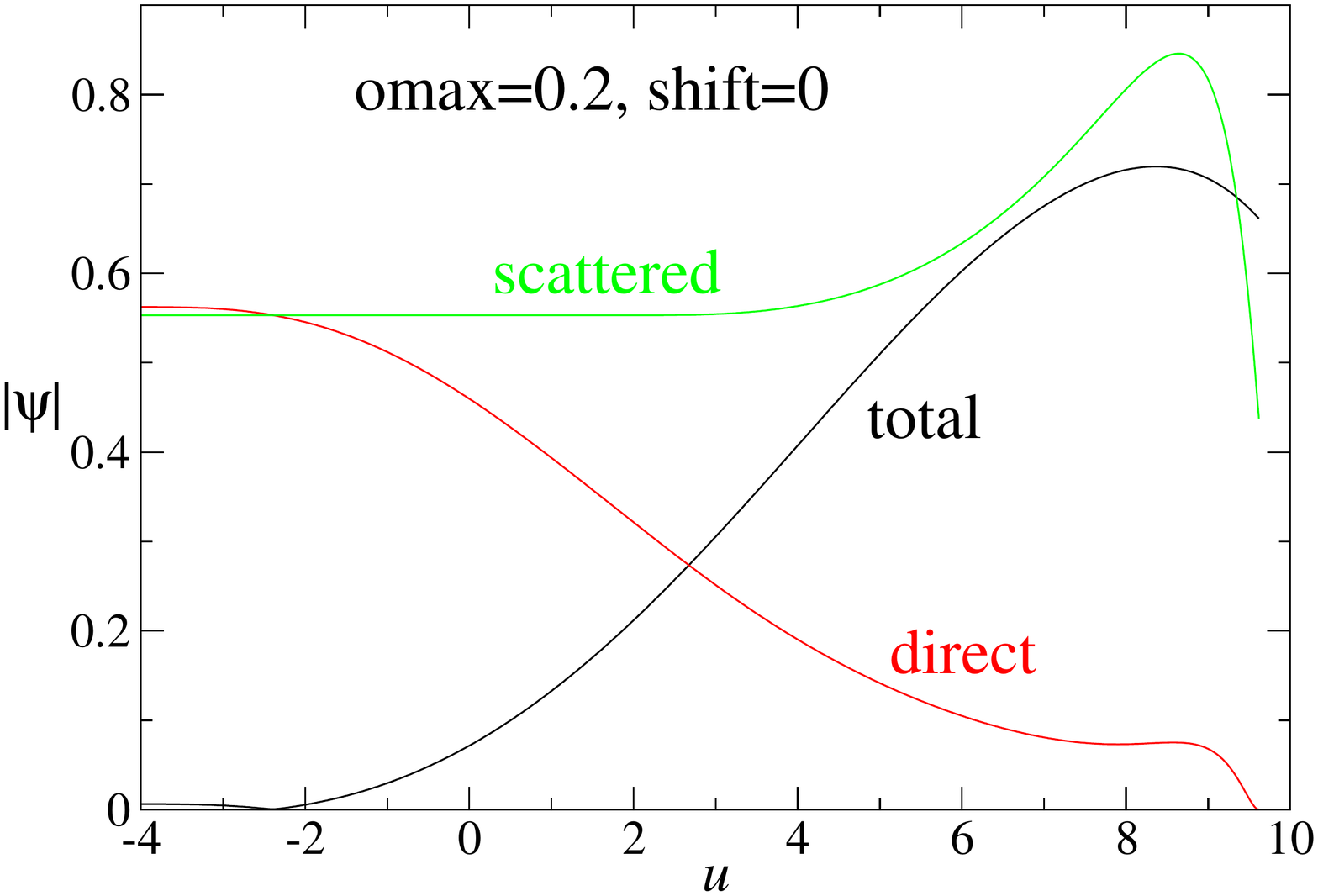}\\
     \includegraphics[width=.45\textwidth ]{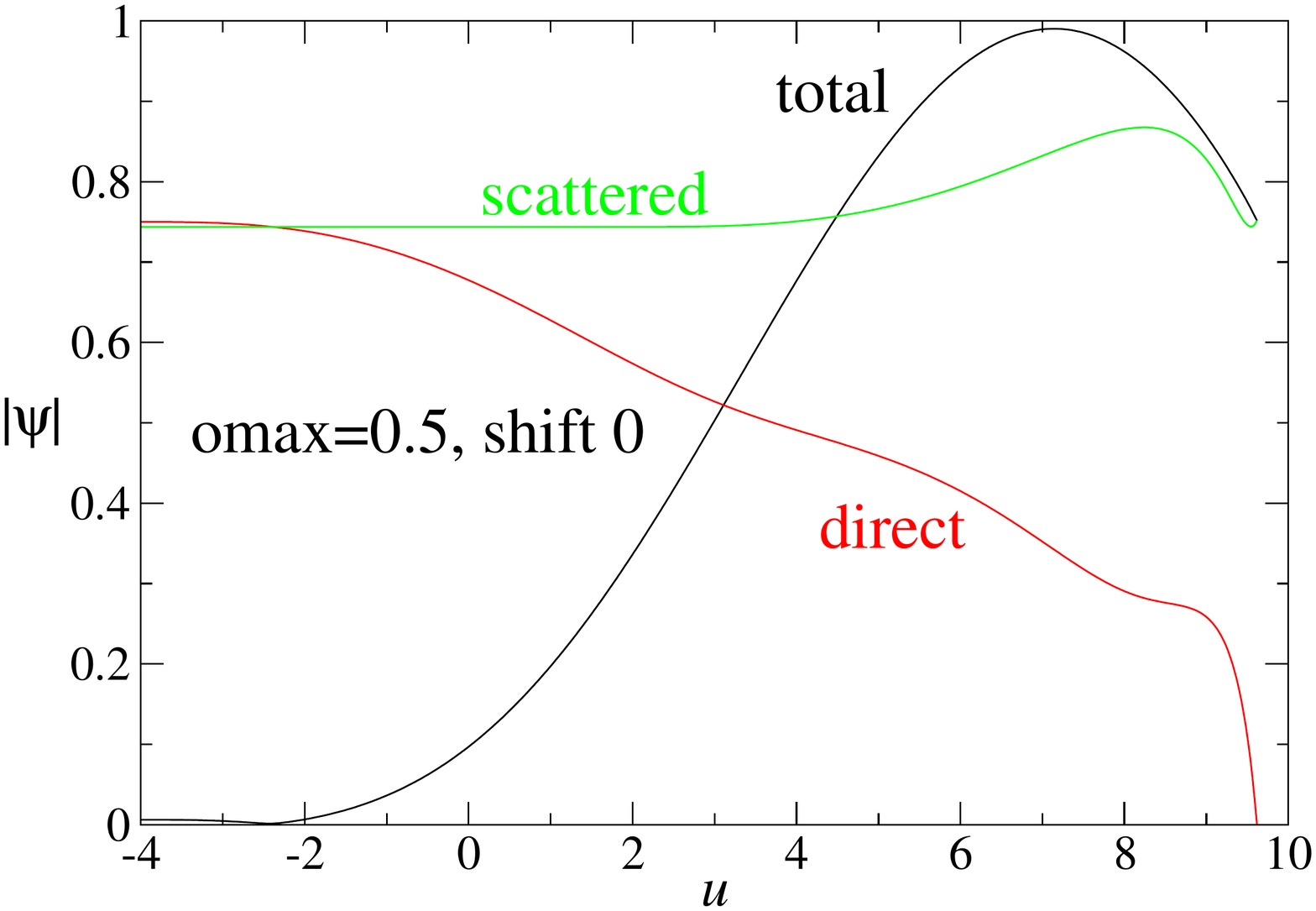}
     \includegraphics[width=.45\textwidth ]{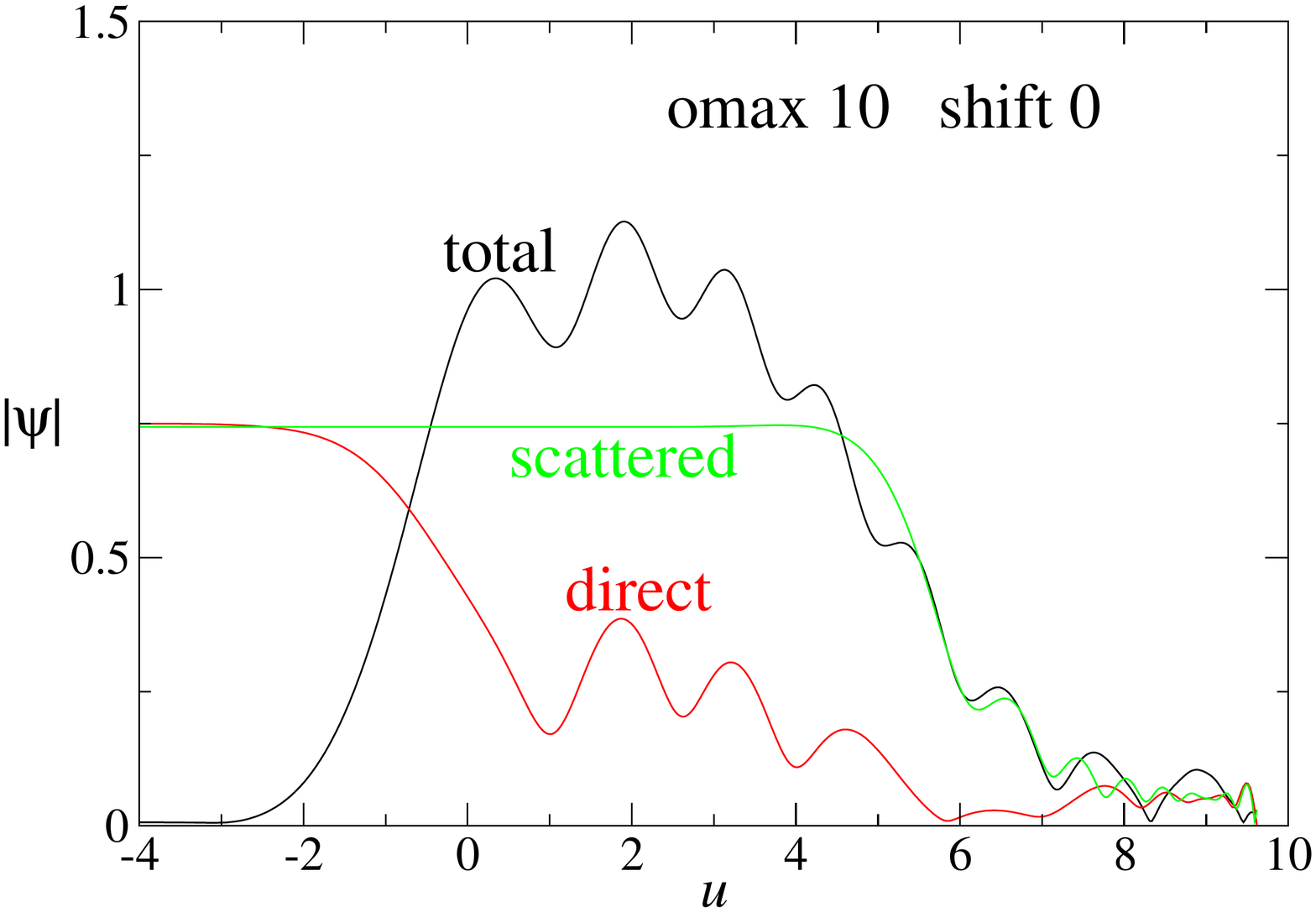}
    \caption{The total radiation complitude, along with the amplitudes 
of the scattered component, and of the direct radiation component,
for several different values
      of $\omega_{\rm max}$.
\label{scatvsdirect}}
  \end{center}\end{figure}

Figure~\ref{intermediate} shows the results for the interesting
intermediate range of $\omax$.  Clearly, as $\omax$ increases the
nature of the curves goes through a transition from the low $\omax$ to
the high $\omax$ curves of Fig.~\ref{extremes}. In these results the
peak amplitude is not correlated with the retarded time of the peak
angular velocity, and is generally earlier.
By exploiting the simplicity of the TDP/FDGF mathematics we can
analyze the source of this behavior. Figure~\ref{scatvsdirect} shows
the radiation amplitudes separately for the direct and the scattered
components of the total, i.e., for the amplitudes computed separately
from the first and second integrals in Eq.~(\ref{eq:psigen}).

The sequence of curves, in Fig.~\ref{scatvsdirect}, from low to high
values of $\omax$ shows interesting trends. At high $\omax$, the
scattered and direct components are in phase at the peak of the total
radiation.  The total amplitude, in fact, is very nearly the sum of
the two individual amplitudes. On the other hand, for low values
$\omax$, the two components of radiation appear to be totally out of
phase.

What is most important to an explanation about the 
results is that in Fig.~\ref{scatvsdirect} the direct contribution 
falls off at late retarded times, times approaching $u_{\rm cross}$, the 
retarded time corresponding to the particle crossing the $x_0$ edge.
The explanation of why the direct radiation falls off approaching
$u_{\rm cross}$ can be extracted from the simplicity of the first
integral in Eq.~(\ref{eq:psigen}), the integral for the direct
radiation.  The definitions in Eq.~(\ref{eq:T1T2}) show that the range
of integration, $T_1-T_2$, shrinks to zero as 
$u\rightarrow u_{\rm
  cross}$. The physical meaning of this is clear in the cartoon in
Fig.~\ref{fig:T1T2} which shows a direct, and a ``bounced'' radial
characteristics from particle to observer. As the particle approaches
the $x_0$ edge, the separation of these two paths shrinks. 
The direct radiation decreases, in fact, because the 
reflected outgoing radiation from the particle tends to cancel the
radiation going directly outward from the particle. As the partice
approaches the reflecting edge at $x_0$, this cancelation tends to be
complete.

There is a general, but not universal tendency for the scattered radiation,
like the direct radiation, also to fall off as 
$u\rightarrow u_{\rm cross}$. What is central, however, to the explanation of the early
amplitude peak is that the direct radiation, an important part of the 
total radiation, must vanish at $u_{\rm
  cross}$, thereby forcing the peak amplitude to occur earlier.

This explanation can be reinforced by a consideration of the relative
phases of the two radiation components.  At the start of motion
($u=-4$ in Fig.~\ref{scatvsdirect}) there can be no radiation; the
particle is starting with no radial or angular acceleration.  It is
interesting that neither the direct nor the scattered component is
zero at the start, but their sum must be, since the total radiation
must vanish.  All curves in Fig.~\ref{scatvsdirect} therefore start
with the two components out of phase, but the subsequent development
is quite different for low and for high $\omax$. For low $\omax$, the
two components remain out of phase. The fact that for low $\omax$ the
amplitude peak occurs at ``late'' times, not much before $u_{\rm cross}$, 
can be ascribed to the vanishing of the subtraction of the out of phase
direct radiation.  (The rise in the scattered radiation at low $\omax$
in Fig.~\ref{scatvsdirect} is not universal; it is not seen in slower
infall models.)  At high $\omax$, the angular modulation of both
components drives them to be in phase. The higher the value of
$\omax$, the more tightly are the two components locked in phase. The
physically required decrease in the direct component as $u$ approaches
$u_{\rm cross}$ along with the observed tendency for the scattered
component also to decrease explains why the amplitude peak at high
$\omax$ is at early times.

It now remains to be seen whether this explanation applies to the
Schwarzschild problem as well as to our TDP toy model.

\section{The EMRI/Schwarzschild Results}\label{sec:Schw}
As pointed out in Sec.~\ref{sec:TDP}, the mathematics of the true
Schwarzschild problem has an obvious similarity to that of the TDP
model, but there are additional reasons to believe that the picture of
the previous section, based on TDP, applies also to Schwarzschild. In
particular, the two integrals in Eq.~(\ref{eq:psigen}) arise from the
poles of the Green function ${\cal G}(x,\widetilde{F};\omega)$. 
The direct integral arises from the $\omega=0$ pole, and
the scattered radiation integral arises from the pole at the QN
frequency. The structure of the Green function for Schwarzschild is
similar, with a pole for the dominant QN mode (as well as for all other QN
modes for a given multipole) and a singularity at $\omega=0$, although
the Schwarzschild Green function has additional
structure~\cite{SchwGF}.

The computation of radiation from the Schwarzschild FDGF is much more
difficult than that for the TDP FDGF, so we take advantage of the
availability of a time-domain evolution code~\cite{PNK}. With that code, 
we repeat the same study as the one performed in the previous sections in 
the context of the TDP model, for the case of gravitational wave (GW) 
emission from EMRIs into a Schwarzschild black hole of mass $M$. Once again, we do
not restrict ourselves to physical inspiral trajectories and instead
use the equivalent of the radial and angular motion used in Sec.~\ref{sec:TDP} and \ref{sec:TDPresults}
with the tortoise coordinate 
\begin{equation}
  r^*=r+2M\ln{(r/2M-1)}\,
\end{equation}
replacing $x$, and the LR location $r^*=1.6317M$ (equivalent to $r=3M$)
in place of $x_0$. Here and throughout we use the notation of the textbook by Misner, 
Thorne and Wheeler~\cite{MTW}, and in particular use $c=G=1$.

\subsection{Model details}
For the radial motion, we use the same ``cubic'' trajectory as was used in the
 TDP model of the previous section, although with different parameter
choices. All models were chosen to have radial motion with $a_0 =
38.6114 M$ and $\tau = 600 M$, and in all cases the particles pass the
LR at time $t = T_{\rm LR} = 348.90 M$, with a radial speed $v_{\rm LR} =
0.3$ (From our previous work~\cite{PNK} we know that these parameters allow
accurate evolutions.). For the angular motion, we use the same law as for the 
TDP models in Eqs.~(\ref{eq:phiofT}).

We had noted in our previous paper~\cite{PNK} that the angular velocity
must have its peak at the light ring. For that reason, our default
choice will be to have $T_{\rm peak}$ correspond to the time ($T_{\rm
  LR}$) at which the particle crosses the LR. We will, however, allow for 
exploration of sensitivity by taking, in analogy with the TDP results, 
$T_{\rm peak}=T_{\rm LR}+\mbox{shift}$, in which we will take shift to 
be $\pm7M$ in addition to $0$. Note that all GW waveforms were extracted at
$r^{*} = 200 M$. Extraction at $r^*=\infty$, equivalent to the ``extraction''
in the TDP model, would make an insignificant difference.

\subsection{Computational results}
\begin{figure}[h]%SCHEXTREME 
  \begin{center}
  \includegraphics[width=.5\textwidth ]{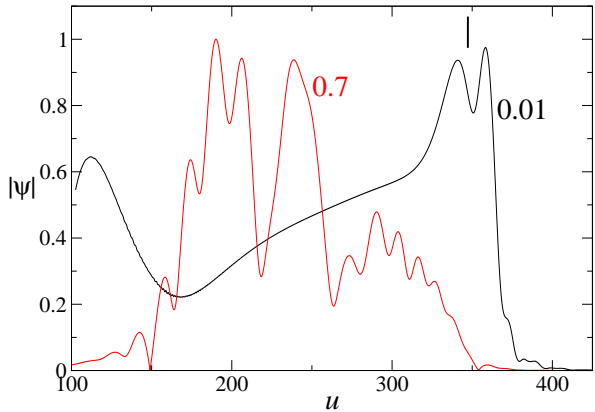}
  \caption{ The (normalized) amplitude for the $\ell,m=2,2$ outgoing GW radiation 
    for the particle trajectories in Schwarzschild spacetime as described in the text. 
    Curves are marked with value of
    $\omax$, the peak angular velocity of the particle, in units of $1/M$. The bold
vertical bar in the upper right of the graph indicates the retarded time corresponding 
to the passage of the particle through the LR. }
  \label{fig:SchExtremes}
  \end{center}
  \end{figure}

We follow the pattern of the previous section and start with the cases
of limiting $\omax$.  In Fig.~\ref{fig:SchExtremes} we show such
limiting cases for shift=0 (i.e., for the maximum of the angular
velocity at the LR). These results show the dominant $\ell,m=2,2$ 
gravitational wave amplitude, for the $M\omax$ values 0.7 and 0.01, as 
indicated. In order to illustrate the peak location most clearly, the amplitude 
dependences on retarded time reported in Fig.~\ref{fig:SchExtremes}, and below in 
Fig.~\ref{fig:SchIntermed}, are normalized to peak amplitude unity.
It is worth noting that gravitational wave amplitudes depend much more on $\omax$, 
than in the TDP case. This strong dependence is due to the strong dependence on $\omega$ 
inherent in the the stress-energy source. Computations, not reported here, with 
a scalar field in the Schwarzschild background, rather than gravitational 
perturbations, show behavior qualitatively similar to the TDP model for the 
dependence of amplitude on $\omax$.

In Fig.~\ref{fig:SchExtremes} the curve of $\omax=0.01/M$ stands also
for all curves of smaller $\omax$. For all such values there is
negligible angular motion during the generation of radiation, and
hence no dependence on $\omax$. In the same spirit, the result for
$\omax=0.7/M$ shows the qualitative nature of the amplitude for high
$\omax$; at larger values of $\omax$, the results have more rapid
oscillations, but the region of the peak amplitude does not change.

A bold vertical bar at retarded time $u=348.9$ indicates the retarded
time corresponding to $T=T_{\rm LR}$ and the maximum of the angular
velocity, for zero shift.  As in the TDP case, the low $\omax$ result
has a peak at approximately  that retarded time while the high
$\omax$ amplitude peak is much earlier than that retarded time.

Figure~\ref{fig:SchIntermed} is the Schwarzschild analog of the TDP
results in Fig.~\ref{intermediate}.  The amplitude as a function of
retarded time is presented for a sequence of intermediate values of
$\omax$, showing the transition from low $\omax$ to high $\omax$. The
qualitative nature of this transition is very similar to that for the
TDP results. (Again, it should be noted that in
Fig.~\ref{fig:SchIntermed}, unlike Fig.~\ref{intermediate}, all peak
amplitudes are normalized to unity.)

  \begin{figure}[h]%SCHINTERMED
  \begin{center}
  \includegraphics[width=.6\textwidth ]{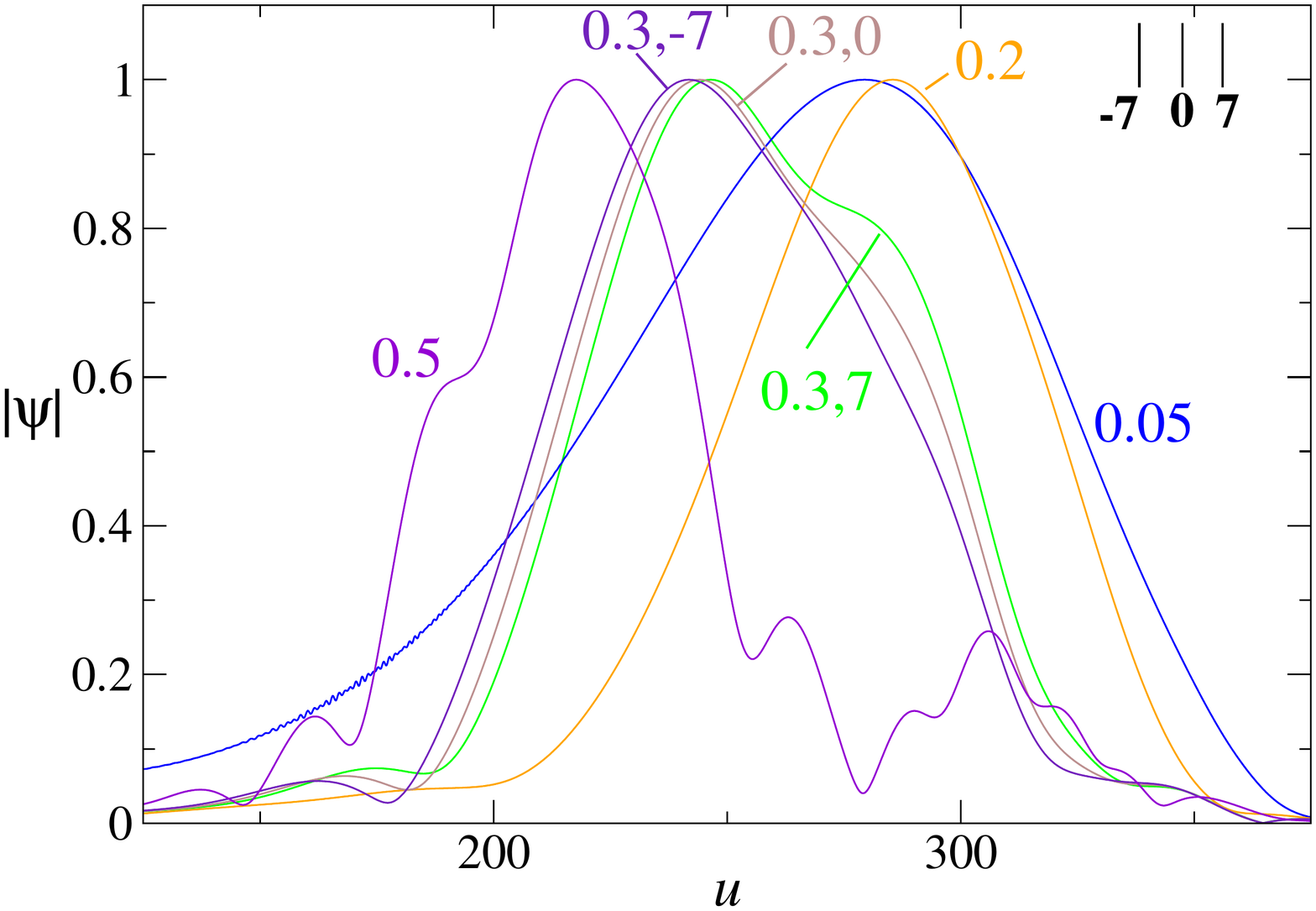}
  \caption{ The (normalized) amplitude of the outgoing radiation for
    the particle trajectories in Schwarzschild spacetime as described in the text. 
    Curves are marked with value of $M\omax$, the peak angular velocity of 
    the particle. For $M\omax=0.3$ the second index in the curve labels
    gives the value of the shift, the time, in units of $M$, by which
    the angular velocity maximum has been shifted. In the upper right
    hand corner, vertical bars indicate the retarded times
    corresponding to the angular velocity maximum for shifts $0,\pm7$
    in units of $M$.  Note that the QN frequency is $0.37$ in units of
    $1/M$. It is clear that only when the value of $\omax$ is
    comparable to the QN frequency, the the GW peaks, shift to earlier
    times.  }
  \label{fig:SchIntermed}
  \end{center}
  \end{figure}

\subsection{Relationship to previously observed GW peak offsets}
As stated before, the fact that the (retarded) peak of the orbital
frequency and the peak of the gravitational radiation do not align has
been known for some time, both in the context of
EMRIs~\cite{BBHKOP,TBKH} and in the comparable-mass binary
systems~\cite{PBBBKPS}. In addition, it has also been noted that
different spin-weighted spherical harmonic modes of the gravitational
waveform peak at different times (see for example, Table II in
Ref.~\cite{BBHKOP} for the Schwarzschild EMRI case). It has also been
known that in the Kerr geometry the radiation peak offset is strongly
dependent on the spin $a/M$ (see Table III in Ref.~\cite{BBHKOP}). In
this section, we argue that the results previously reported about the 
peak misalignment are consistent with the central observations  in the  
current work.

First, consider the case of gravitational wave emission from particle
motion in Schwarzschild spacetime. Examining the dependence of the
peak offset for fixed $\ell$ modes (in particular, $\ell=2$) in Table
II of Ref.~\cite{BBHKOP} suggests that the higher $m$-modes peak at
earlier times. This is exactly what we have pointed out in the
previous sections of this paper. In our TDP model, we changed the
value of the angular frequency instead, but note that is
mathematically equivalent to changing the $m$-mode, of course. (It is
important to fix the value of $\ell$ for a proper comparison with our
current work, in order for the potential function to stay unchanged.)

In the Kerr context, Table III in Ref.~\cite{BBHKOP} depicts the peak
offset of the $\ell=m=2$ mode as a function of the Kerr parameter
$a/M$. It is clear from those data that as $a/M$ increases the
gravitational waveform peaks earlier.  Once again, this is consistent
with the expectations from the previous sections in this work;
a higher value of $a/M$ implies higher orbital frequency (a smaller
ISCO allows the particle to orbit a lot closer to the Kerr hole) and
those cases peak at earlier times. 
Of course, 
the Kerr potential is strongly spin-dependent, and that likely
plays a role in the timing of the peaks as well. Moreover for the
high-spin cases, while the peak offset is large, it is also true 
that the peak ``flattens out'' considerably, making it harder to 
pinpoint its precise location. 

%%-----------------Sat morn-------------------
As examples of these offsets we present results adapted from
Refs.~\cite{BBHKOP,TBKH}.  These results were generated using a
high-accuracy and high-performance Teukolsky equation solver in the
time-domain, with a point-particle source term~\cite{mycode}.
Table~\ref{Kerrtable} shows the delay in the dominant ${\ell,m}=2,2$
mode of gravitational radiation for a particle spiralling in on an
equatorial orbit into a Kerr black hole. The value of $a/M$ indicates
the spin of the Kerr hole; in the first column, a negative $a/M$
indicates retrograde inspiral. The second column shows the delay between 
the moment the radiation peaks, and the moment when the radiation arrives 
from the peak of the angular velocity. For all but the retrograde inspiral, 
this delay is negative; the peak of the amplitude is earlier than the signal 
coming from the peak of angular velocity.

This table also shows a comparison of frequencies that are
relevant to our explanation of this delay. The third column shows the peak 
of the angular velocity $d\phi/dt$. The angular velocity is
given in units of $1/M$ and multiplied by 2 since the 2,2 radiation
produced by angular motion should have frequency twice that of the
angular motion. The fourth column, with the quasinormal frequencies for
the least damped 2,2 mode shows that this frequency is close to the
angular velocity and thus the mechanism for the peak offset is likely 
to be the same that we have presented in this current work. 

\section{Conclusion}\label{sec:conc}

Though peak time offsets have been important in waveform construction,
all that has been known is that such a time offset exists and that it
is different (a) in the Schwarzschild background for gravitational
waves in different multipole modes, and also (b) in the Kerr background
for even a given multipole mode, but different values of $a/M$.  We know of 
no claim in the literature that suggests an explanation of this offset, what 
it depends on, or whether (a) and (b) are related.

In this paper we present insights that provide at least the beginning
of an explanation of the peak offset phenomenon. These insights are
the results of exploiting the simplicity of a model first explored in
our previous paper~\cite{PNK}. In particular, (i)~the curvature
potential of the Schwarzschild EMRI problem is replaced by a
simplified truncated dipole potential; (ii)~the source is understood
through the Fourier domain Green function; (iii)~we do not restrict
ourselves to geodesic orbits, but use trajectories that allow us to
probe the mechanisms for the generation of radiation.

This simplicity has produced a very simple explanation of why the
amplitude peak for low $\omax$ is at the ``expected'' late retarded
time, but for high $\omax$ is significantly earlier. In simplest terms,
the idea is that the radiation consists of two components, direct
radiation and scattered radiation, and that the direct component
vanishes at the ``expected'' retarded time. For low $\omax$ the two
components are out of phase, so the late-time vanishing of the
subtracted direct radiation means that the peak, dominated by the
scattered radiation, can occur at late times. For high $\omax$ the two
components are in phase. The vanishing of the late-time direct
radiation, therefore, removes its addition to the total, and pushes
the peak total to earlier times.

The separation of the direct and scattered components of the radiation
for the Schwarzschild EMRI problem is not simple, but the qualitative
similarity of our results for the TDP and for the Schwarzschild
background, both for gravitational waves and for scalar waves (not
reported here), makes a strong case that the same mechanism is at
work. The case is particularly strong for the claim that the
replacement of the Schwarzschild curvature potential by the TDP
potential is justified in seeking an explanation.

Further support comes from the details of the Schwarzschild results in
the literature.  In Table II of the EMRI studies reported for the
Schwarzschild background in Ref.~\cite{BBHKOP}, the time by which the
amplitude peak precedes the angular velocity peak, for a given $\ell$
mode, increases with the $m$ index of the mode. This increase is
mathematically equivalent to increasing the angular velocity. The
reported results, then, show that as the angular velocity increases,
the peak moves earlier, as in the results we report above, both for
the TDP and Schwarzschild models.

Turning to the Kerr case, in Table \ref{Kerrtable} we present results 
\begin{table}
  \begin{tabular}{|c|c|c|c|}
\hline
$a/M$&   $h_{22}$\ offset&   $2M\omega_{\rm peak}$&  $M\omega_{\rm QN}$     \\
\hline
-0.9 &         +2.0   &      0.18   &            0.30    \\
0.0   &      -3.0     &     0.28     &          0.37     \\
0.5   &      -7.0     &     0.38     &          0.47    \\
0.9   &      -40.0    &     0.64     &          0.67     \\
0.99  &      -60      &     0.84      &         0.87         \\
\hline 
        \end{tabular}
\caption{The time offsets of the $\ell,m=2,2$ peak of radiation from particles
  spiralling into Kerr black holes with spin parameter $a$, and mass
  $M$.  The second column shows the offset (in units of $M$) by which the GW 
  peak follows the signal from the maximum of the angular velocity.  The third column
  shows twice the peak orbital frequency (in units of $1/M$) of the
  inspiralling particle. The fourth column is the quasinormal frequency of the 
  dominant (least damped) $\ell,m=2,2$ mode. It is clear that when the last two
  columns have values that are comparable, a significant offset is observed. \label{Kerrtable}}
  \end{table}
of EMRI studies adapted from Refs.~\cite{BBHKOP,TBKH}. Of note is the
fact that the peak frequency is at a value quite close to that of the
real part of the least damped quasinormal frequency for the GW 2,2
mode. These frequencies correspond to the high/intermediate
frequencies of Secs.~\ref{sec:TDPresults} and \ref{sec:Schw} and this
strongly suggests that the same mechanism may be at play in the Kerr
case as well.

It remains to be seen whether the strong $a/M$ dependence of the
offsets in the Kerr geometry can be fully explained in terms of the
same mechanism as in the TDP and Schwarzschild cases. The insights
about these peak offsets, as presented in this paper, can most
effectively be used in the EOB approach to waveform construction.

\section{Acknowledgments} We are greatly indebted to Alessandra Buonanno for her 
thorough review of an earlier version of this manuscript. Her detailed
feedback has significantly improved this current version. We would
also like to acknowledge discussions and helpful remarks made by Scott
Field, Scott Hughes, and Andrea Taracchini in the context of this work.
G.K. acknowledges research support from NSF Grants No. PHY-1303724 and
No. PHY-1414440, and from the U.S. Air Force agreement
No. 10-RI-CRADA-09.

\pagebreak


\begin{thebibliography}{11}
\bibitem{radial} Since there is no physical model connected to the TDP
model, the term ``radial'' is not strictly meaningful, but is  convenient.
\bibitem{explainT2} The presence of $T_1$ in the first integral might
  seem to suggest that it does not quite describe direct radiation,
  because $T_1$ depends on $x_0$. In the details of the derivation,
  however, it turns out that at early times, the particle influence
  propagates inward, reflects off the $x_0$ edge, reverses its sign,
  and propagates outward. At particle times earlier than $T_2$, the
  directly outgoing influence and the influence reflected off the edge
  cancel.
\bibitem{scaling} The scale in the TDP problem is set by $x_0$, which
  can be thought of as playing a role analogous to that of the mass
  parameter in the Schwarzschild spacetime. In principle, we could
  express all parameters in dimensionless form, for example, we could
  use $\tau/x_0$ in place of $\tau$.  Since we are choosing $x_0=1$,
  all parameters can be thought of as being de-dimensionalized in this
  way.
\bibitem{SchwGF} There are an infinite number of QN frequencies for the Schwarzschild
spacetime, and hence an infinite number of  corresponding poles in 
the Schwarzschild GF. In addition, the Schwarzschild GF has a branch point
at $\omega=0$.
\end{thebibliography}
\end{document}